\documentclass[%
 reprint,
 amsmath,amssymb,
 aps,
]{revtex4-2}
\usepackage[british,UKenglish,english]{babel}
\usepackage{listings}
\usepackage{color}
\usepackage{xcolor}
\usepackage{systeme}
\usepackage{listings}
\usepackage{amsmath}
\usepackage{comment}
\usepackage{leftidx}
\usepackage{amssymb}
\usepackage{cases}
\usepackage{mathtools}
\usepackage[toc,page]{appendix}
\usepackage{systeme}
\usepackage{bm,bbm}
\usepackage{natbib}
\usepackage{upgreek}
\usepackage{enumerate}
\usepackage{url}
\usepackage{float}
\usepackage{graphicx}
\usepackage{dcolumn}
\usepackage{bm}

\newcommand{\p}{\partial}
\newcommand{\lan} {\langle}
\newcommand{\ran} {\rangle}
\renewcommand{\d} {\mathrm{d}}
\newcommand{\blue}  [1]{\textcolor{black}{#1}}

\newcommand{\purple}  [1]{\textcolor{black}{#1}}

\begin{document}

\preprint{APS/123-QED}

\title{Expressing turbulent kinetic energy as coarse-grained enstrophy or strain deformations} 

\author{Damiano Capocci}
 \email{dcapocci@ac.uk.ed}
\affiliation{
Department of Physics, University of Rome Tor Vergata
}\thanks{{Currently employed at School of Physics and Astronomy, The University of Edinburgh, Edinburgh, United Kingdom}}

\begin{abstract}
\noindent In turbulent flows, the \blue{fluid element} gets deformed by chaotic motion due to the formation of sharp velocity gradients. A direct connection between the element of fluid stresses and the energy balance still remains elusive. 
Here, an exact identity of incompressible turbulence is derived linking the velocity gradient norm across the scales with the total kinetic energy. In the context of three-dimensional (3D) homogeneous turbulence, this relation can be specialised obtaining the expression of total kinetic energy decomposed either in terms of deformations due to strain motion or via the resolved-scale enstrophy of the fluid element.
Applied to data from direct numerical simulations (DNS) \blue{describing homogeneous and isotropic turbulence}, the decomposition reveals that, beyond the scales dominated by the external forcing, contractile and extensional deformations account approximately for $55\%$ and $40\%$ of the kinetic energy of the associated scale while less than the remaining $5\%$ is carried by the indefinite-type stresses. 
From these two identities one can derive an exact expression for the kinetic energy spectrum \blue{which is based solely on real space quantities} providing a characterisation of the Kolomogorov constant as well. Numerical evidences \blue{show} that this formulation of the energy spectrum reproduces the power-law behaviour of the Kolmogorov spectral scaling.

\end{abstract}
\maketitle

Among the countless scientific contexts that deal with turbulent flows, two characteristics are recurrent: the nonlinearity and the multiscale nature of the velocity field leading to a spatio-temporal chaotic dynamics. Turbulence is a common emerging property in water or air when the value of the viscosity is \emph{small} compared to the related typical sizes and velocities. A turbulent flow is characterised by the formation of small-scale vortical motion from a large scale injection of energy; this phenomenon is usually referred to as the \emph{cascade} of energy across the scales. This energy transfer is responsible of the production of velocity gradient (VG) at every scale which, in turn, determines the deformations that the element of fluid undergoes. \blue{Characterising the interplay between turbulent kinetic energy and flow deformation not only pertains to fundamental knowledge but is also relevant for applications ranging from geophysics \cite{buaria2019,thorpe2005,pedlosky2013} to astrophysics \cite{cho2003,goldstein1995,goldreich1997}}. 
\blue{To the author's knowledge, the functional dependence between the total kinetic energy and both the intensity of strain and rotational motions is still unknown}.

In this work, by proving a more general identity, two expressions for the total kinetic energy are derived in homogeneous turbulence: one written in terms of the rate of strain deformation and the other based on the squared vorticity, commonly referred as \emph{enstrophy}. As an extension of the latter, a corresponding exact identity for the kinetic energy spectrum is derived through which it is possible to connect the \blue{kinetic energy spectrum} with filtered enstrophy. The derived identities are verified via data from direct numerical simulation (DNS) and used to reveal the percentages of the purely contractile, the purely extensional and the \emph{indefinite-type} stresses that constitute the kinetic energy across the scales. The validity of the novel representation of the kinetic energy spectrum is tested as well.

The Navier-Stokes equation describing the evolution of an incompressible velocity field $\bm{u}(\bm{x},t)$ is the following:
\begin{equation}
\partial_t u_i = - u_j \dfrac{\p u_i}{\p x_j}  - \dfrac{1}{\rho} \dfrac{\p p}{\p x_i} + \nu \nabla^2 u_i + f_i
\label{eq:NS}
\end{equation}
where $\nu$ is the kinematic viscosity and $\rho$ the mass density. The linear \emph{viscous term} $\nu \nabla^2 u_i$ corresponds to the dissipation, being related to the conversion of kinetic energy into heat. The pressure field $p(\bm{x},t)$ enforces the incompressibility constraint $\partial_i u_i = 0$ and $f_i$ is a generic external force injecting energy into the system. The VG tensor
$u_{i,j}=\partial u_i / \partial x_j $ describes the flow geometry in
terms of strain rate $S_{ij} = (u_{i,j} + u_{j,i})/2$ and rotation rate
$\Omega_{ij} = (u_{i,j} - u_{j,i})/2$, which can also be expressed as the vorticity $\omega_i = \epsilon_{ijk} \Omega_{jk}$. Furthermore, its Frobenius norm determines the rate at which kinetic energy is dissipated into thermal energy in $\varepsilon= \nu \, u_{i,j} u_{i,j}$. Eq.~\eqref{eq:NS} admits one dimensionless control parameter which is the Reynolds number $Re_L=U\,L/\nu$, where $U$ and $L$ are characteristic velocity and length scale of the flow, respectively.  

By applying a low-pass filter $G^{\ell}$, we can separate the large-scale dynamics from the small-scale ones \citep{germano1992}:
\begin{equation}
\overline{u}^\ell_{i} = G^\ell * u_i, \quad \quad \tilde{\overline{u}}^\ell_i = {\tilde{G}^\ell} \cdot \tilde{u}_i 
\label{eq:fourier} 
\end{equation} 
where $\tilde{( \cdot )}$ indicates the Fourier transform and $*$ the convolution product. As regards the properties of $G^\ell$, we require \blue{it to be} an even function with volume average equal to unity \cite{pope2001}.

 The application of a low-pass filter to eq.~\eqref{eq:NS} provides:
\begin{equation}
\partial_t \overline{u}^\ell_i = - \overline{u}^\ell_j \, \dfrac{\p \overline{u}_i^\ell}{\p x_j}  - \dfrac{1}{\rho} \dfrac{\p \overline{p}^\ell}{\p x_i} + \nu \nabla^2 \overline{u}_i^\ell + \overline{f}^\ell_i - \p_j \tau^\ell_{ij}
\label{eq:NS_filt}
\end{equation}
where $\tau_{ij}^\ell = \overline{u_i u_j}^\ell - \overline{u}^\ell_i \overline{u}^\ell_j$ is the so-called subgrid-scale stress (SGS) tensor, representing the effective stress due to the features smaller than $\ell$ on the resolved-scale velocity. Therefore, we can define the kinetic energy related to scales larger than $\ell$ i.e. the \emph{resolved-scale energy} as $E_K^\ell(\bm{x},t) = \frac{1}{2} \overline{u}^\ell_i (\bm{x},t) \, \overline{u}^\ell_i (\bm{x},t)$ and the kinetic energy at scales smaller than $\ell$ \purple{which is the trace} $\frac{1}{2} \tau^\ell_{ii}(\bm{x},t)$ as the \emph{unresolved-scale energy}. Note that, by definition, the sum of these two quantities determines the filtered kinetic energy:
\begin{equation}
\dfrac{1}{2} \overline{u_i \, u_i}^\ell =  E^\ell_K  + \dfrac{1}{2} \tau^\ell_{ii}
\label{eq:total_en}
\end{equation} 
where we omit the spatio-temporal dependence. If the filter kernel is non-negative i.e. $G^\ell(\bm{r}) \ge 0$ $\forall \bm{r} $, then $\tau_{ii}(\bm{x}) \ge 0$ $\forall \bm{x}$, hence $\tau_{ii}$ can be correctly interpreted as an energy \citep{vreman1994}. We also remind that $\overline{u}^{\ell=0}(\bm{x},t) = u(\bm{x},t) $ implying that $\tau^{\ell=0}=0$ and $E_K^{\ell=0} \equiv E_K = \frac{1}{2} {u}_i (\bm{x},t) \, {u}_i (\bm{x},t) $ \blue{which is the kinetic energy}. 

At this stage, we note that, if a scalar function $g$ admits the Fourier representation \footnote{Note that, for an unbounded domain, the Fourier representation can be bypassed by requiring that $g(\bm{x})$ goes to zero as $||\bm{x}|| \to \infty$. The bypass is formal in periodic boundaries which in turn would imply the expansion in Fourier series. The same considerations applies to the derivation of eq.~\eqref{eq:perrys_meth} in case one wants to avoid the Fourier expansion.}, it is straightforward to conclude that the relation between the volume averages $\lan \overline{g}^\ell \ran = \lan g \ran$ is a consequence of the properties of the (generic) filter kernel. This observation becomes relevant for the calculation of the volume average of \eqref{eq:total_en}. Indeed, as already noted by \cite{aluie2017}, we can neglect the filter application on the LHS obtaining:
\begin{equation}
\lan E_K \ran =  \lan E^\ell_K \ran + \dfrac{1}{2} \lan \tau^\ell_{ii} \ran
\label{eq:subtrac}
\end{equation}
where the filtering parameter, belonging only to the RHS, \blue{broadly} resembles a \emph{subtraction point} of the renormalization theory \cite{zinn2010}. 

At this point, in order to satisfy the non-negativity of the filter kernel, we center our discussion on the Gaussian kernel:
\begin{equation}
G^\ell(\bm{r}) = \left(2\pi \ell^2 \right)^{-3/2} e^{-\bm{r}^2/2\ell^2}, \quad \quad {\tilde{G}^\ell}(\bm{k}) = e^{-\bm{k}^2\ell^2/2}
\label{eq:gaussian_filt}
\end{equation}
As a specific feature of the Gaussian filter, there exists an exact decomposition of the SGS stresses \cite{johnson2020}\cite{johnson2021} that reads:
\begin{equation}
\tau_{ik}^\ell = \int_{0}^{\ell^2} \hspace{-0.1cm}  \, \d \theta \, \overline{ \, \overline{u}^{\sqrt{\theta}}_{i,j} \, \overline{u}^{\sqrt{\theta}}_{k,j} }^{\sqrt{\ell^2 - \theta}}
\label{eq:perrys_meth}
\end{equation}
expressing SGS stresses \footnote{Note that the methodology of \cite{johnson2020} requires homogeneity or periodic boundaries in order to set to zero the boundary terms and not for the derivation of the decomposition \eqref{eq:perrys_meth} itself. } as the sum of the contributions of VG fields from the scales $\leq \ell$ that are in turn projected into the larger scales by the complementary filter $\sqrt{\ell^2 - \theta}$. It follows that the trace of \eqref{eq:perrys_meth} allows us to recast eq.~\eqref{eq:subtrac} into:
\begin{equation}
\lan E_K \ran = \lan E^\ell_K \ran + \dfrac{1}{2} \int_{0}^{\ell^2} \hspace{-0.1cm}  \, \d \theta \, \lan \overline{u}^{\sqrt{\theta}}_{i,j} \, \overline{u}^{\sqrt{\theta}}_{i,j} \ran  
\label{eq:eq_interm}
\end{equation}
where the filtering at $\sqrt{\ell^2 - \theta}$ in the averaged trace of \eqref{eq:perrys_meth} can be ignored by virtue of the filter-invariance property of scalar functions mentioned above. In addition, the exchange of the volume averaging and the integral over the scales is granted by the assumption that \blue{the} VG \blue{norm is} bounded. \blue{By assuming finite energy, even if} we let our system to be characterised by \blue{infinite length scales}, \blue{it} would immediately imply that $\displaystyle \lim_{\ell \to \infty} \lan E^\ell_K \ran = 0$. The prescription imposed by this limit appears as a reasonable assumption for any classical physics system which is \emph{well-defined} in its range of scales \footnote{The length contraction in general relativity would already pose conceptual difficulties to the present filtering approach.}. A detailed and more generalised analysis on both space and scale convergence of \eqref{eq:eq_interm} will be provided elsewhere \footnote{However, in the applications, the interest \blue{clearly} focuses on systems characterised by a maximum allowed scale for which the integration upper bound of \eqref{eq:perrys_meth} does not diverge ensuring a rigorous exchange of integrals to derive \eqref{eq:eq_interm}.}.  
By applying the previous considerations and calculating the limit for infinite filtering-scale on the RHS of \eqref{eq:eq_interm}, we get:
\begin{equation}
\lan E_K \ran =  \dfrac{1}{2} \int_{0}^{\infty} \hspace{-0.1cm}  \, \d \theta \, \lan \overline{u}^{\sqrt{\theta}}_{i,j} \, \overline{u}^{\sqrt{\theta}}_{i,j} \ran  
\label{eq:eq_almost_final}
\end{equation}
that can be eventually combined with \eqref{eq:eq_interm} obtaining the more general expression:
\begin{equation}
\lan E_K^\ell \ran =  \dfrac{1}{2} \int_{\ell^2}^{\infty} \hspace{-0.1cm}  \, \d \theta \, \lan \overline{u}^{\sqrt{\theta}}_{i,j} \, \overline{u}^{\sqrt{\theta}}_{i,j} \ran.  
\label{eq:eq_almost_final_ell}
\end{equation}
Eqs.~\eqref{eq:eq_almost_final} and the resolved-scale counterpart \eqref{eq:eq_almost_final_ell} can be conceived as an exact gradient-tensor decomposition of kinetic energy. The above equations also go beyond the standard representation of kinetic energy as the squared Euclidean norm of velocity presenting the mean energy budget as the sum over the scales of the squared VG tensor norm. 

At this point, in order to quantify eq.~\eqref{eq:eq_almost_final_ell} we consider data from DNS. Our database \citep{damianinoHDdata} describes stationary homogeneous and isotropic turbulence where eq.~\eqref{eq:NS} is evolved in a triply-periodic box with a stochastic forcing active in the wavenumber band $k \in [0.5,2.4]$ with $k=\pi/\ell$, whose midpoint $k_f$ determines the characteristic forcing scale $L_f$. This dataset is marked by a $Re_\lambda = 327$ where $\lambda = E_K\,\sqrt{15 \nu/ \langle \varepsilon \rangle}$ is the Taylor-scale.

\begin{figure}
	\begin{center}
         \includegraphics[width=1.1\columnwidth]{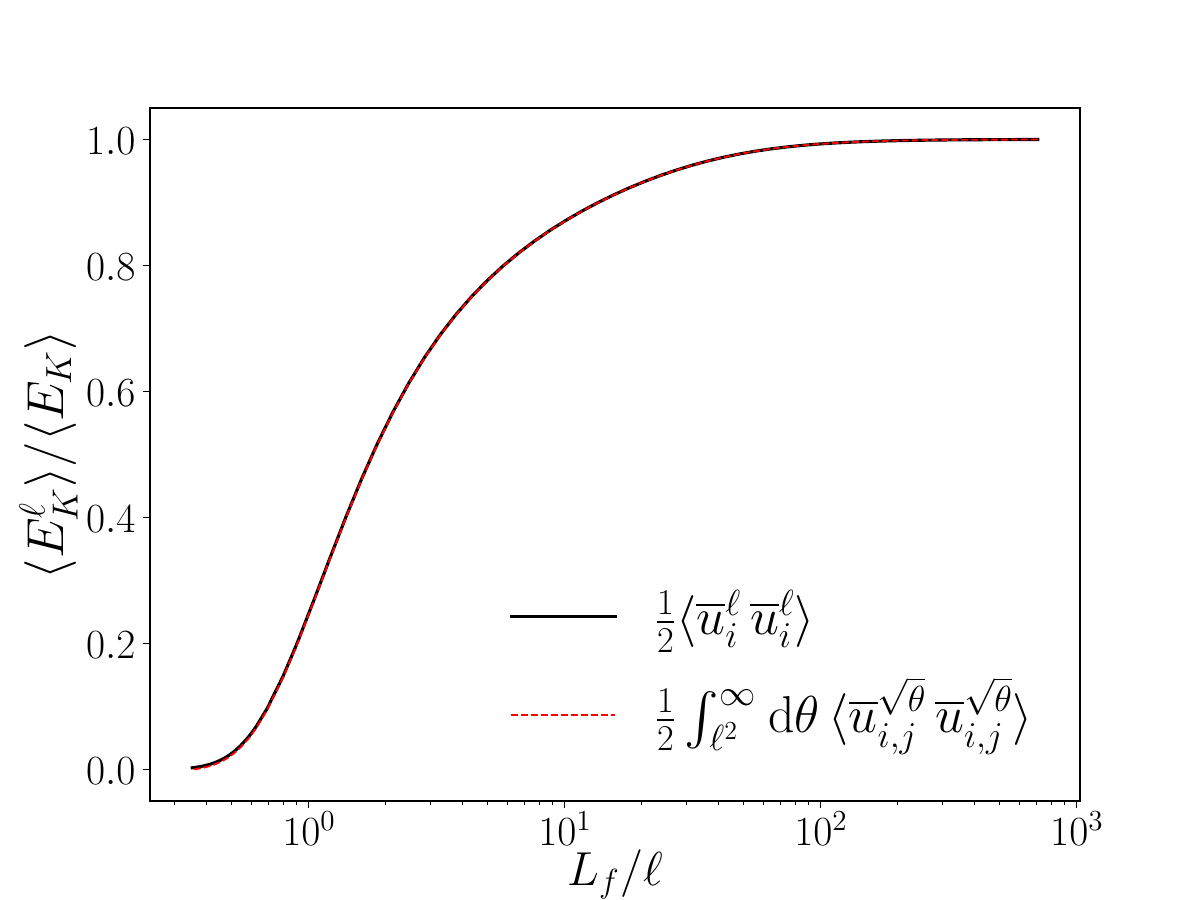} 
        \end{center}
	 \caption{Resolved scale mean kinetic energy normalised by the unfiltered value ${\lan E^\ell_K \ran}/{ \lan E_K \ran}$ as a function of the dimensionless parameter $L_f/\ell$ calculated from the standard definition and from eq.~\eqref{eq:eq_almost_final_ell}. The two curves clearly overlap.} 
	 \label{fig:energy_filt_ratio}
\end{figure}

In the following quantification, we are relying on one instantaneous configuration of the velocity field. In particular, fig.~\ref{fig:energy_filt_ratio} compares \eqref{eq:eq_almost_final_ell} with the \emph{usual} expression of the averaged resolved-scale kinetic energy. Both these two expressions are normalised by the total kinetic energy and displayed as a function of the adimensional parameter $L_f/\ell$. In consequence of the validity of \eqref{eq:eq_almost_final_ell}, we clearly observe that the two displayed profiles are indistinguishable.

Apart from the veracity of the above identities, fig.~\ref{fig:energy_filt_ratio} highlights that about $80\%$ of the total energy budget is contributed by \emph{middle}- and \emph{large}-scale structures of size $\ell \geq L_f/10$. Capturing the remaining $20\%$ requires including nearly two orders of magnitude of smaller scales. While VG tensors are peaked at \emph{small} scales \cite{tennekes1972}, it is crucial to emphasize that their contributions to the kinetic energy balance in eq.~\eqref{eq:eq_almost_final} are strongly suppressed by the squared length scale integration which, in turn, empowers the weaker contributions of gradients arising from \emph{middle}- and \emph{large}-scale structures.

As a conclusion of this \purple{preliminary} analysis, it is worth to underline that in homogeneous turbulence, the first Betchov relation \cite{betchov1956}
\begin{equation}
\lan \overline{S}^\ell_{ij} \overline{S}^\ell_{ij} \ran = \dfrac{1}{2} \lan \, || \overline{\bm{\omega}}^{\ell} ||^2 \, \ran ,
\label{eq:first_betchov}
\end{equation} 
holding $\forall \ell$, implies that the mean (filtered) VG norm can be either written in terms of the (filtered) squared strain or via the (filtered) enstrophy.

\section{Mean kinetic energy in terms of fluid element deformations}

In this section, as introduced above, we are going to express the mean squared VG appearing on the RHS of \eqref{eq:eq_almost_final_ell} only in terms of the resolved-scale strain-rate squared via the application of \eqref{eq:first_betchov}. \blue{Additionally, since} \eqref{eq:first_betchov} is an identity involving kinematic invariants, we can choose the reference frame where $S_{ij}$ becomes diagonal, the so-called \emph{principal axis frame}. We also remark that the strain-rate tensor, being symmetric, has three real eigenvalues $\lambda_i$ with $i \in \{1,2,3\}$ related to a set of orthogonal eigenvectors. Moreover incompressibility imposes the pointwise constraint $\overline{\lambda}^\ell_1 + \overline{\lambda}^\ell_2 + \overline{\lambda}^\ell_3 = 0$. As a consequence, the smallest eigenvalue $\overline{\lambda}^\ell_1 \leq 0$ is always contractile along the direction of its eigenvector and the largest one $\overline{\lambda}^\ell_3 \geq 0$ is always related to an extensional direction while the intermediate $\overline{\lambda}^\ell_2$ can be either positive or negative from which the definition \emph{indefinite-type}. \blue{In light of the preceding discussion}, we express \eqref{eq:eq_almost_final_ell} via the (sorted) strain-rate tensor eigenvalues obtaining:
\begin{align}
\lan E_K^\ell \ran =&  \int_{\ell^2}^{\infty} \hspace{-0.1cm}  \, \d \theta \, \lan \, (\overline{\lambda}^{\sqrt{\theta}}_1)^2 + (\overline{\lambda}^{\sqrt{\theta}}_2)^2 + (\overline{\lambda}^{\sqrt{\theta}}_3)^2 \, \ran \label{eq:eq_final_ell_eig} \\
=& \, P_1^\ell + P_2^\ell + P_3^\ell  \,.  
\label{eq:eq_final_ell_eig_2}
\end{align}
where each $P_i^\ell$ represents the contribution of the $i$-th strain-rate eigenvalue to the mean resolved-scale kinetic energy. \blue{The above equation describes the energy budget via the intensity of stresses along their principal axes. }

\begin{figure}
	\begin{center}
         \includegraphics[width=1.1\columnwidth]{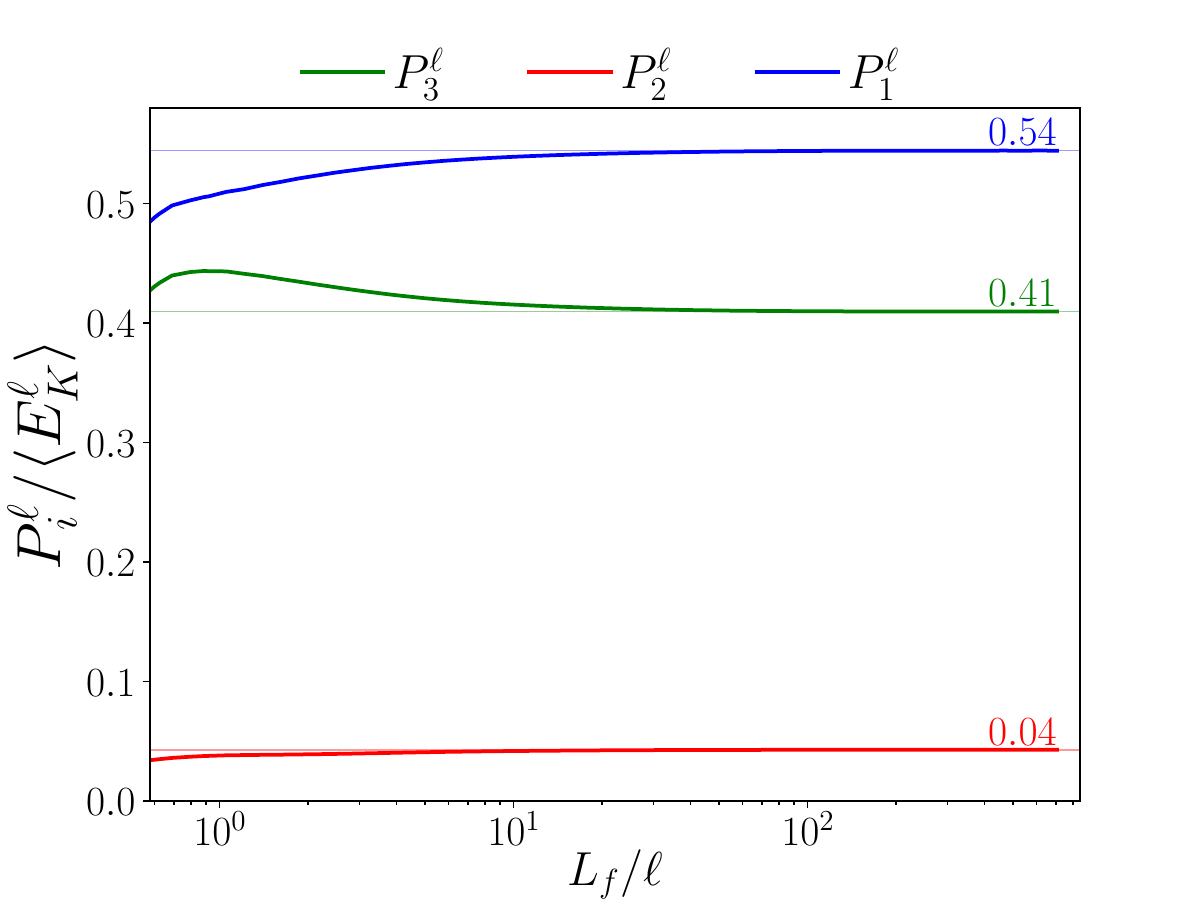} 
        \end{center}
	 \caption{Strain-rate eigenvalues contributions $P_i^\ell$ normalised by the resolved-scale as a function of $L_f/\ell$. Both the axes show dimensionless quantities. The values of the curves sum to unity scale by scale. The displayed percentages refer to the smallest resolved-scale. The horizontal thin lines, with their corresponding $y$-axis (rounded) values, are added manually to emphasize the values at which $P_i^\ell$ saturates.}
	 \label{fig:P_ratios}
\end{figure}

By using the abovementioned data, in eq.~\eqref{eq:eq_final_ell_eig} we can quantify the role of each $P_i^\ell$ across the scales. In fact, fig.~\ref{fig:P_ratios} displays the adimensional $P_i^\ell$ normalised by $\lan E^\ell_K \ran$ as a function of the non-dimensional parameter $L_f/\ell$ \blue{describing} the percentage of each $P_i^\ell$ in the kinetic energy balance at the scale $\ell$. We first notice that, apart from the region $L_f/\ell \lesssim 1$ where the flow geometry is governed by the external forcing, each $P_i^\ell$ is monotonic where $\overline{\lambda}^\ell_1$ increases, $\overline{\lambda}^\ell_2$ \blue{weakly increases, being \blue{essentially} scale-invariant,} while $\overline{\lambda}^\ell_3$ decreases. As we progress towards the \emph{small} scales each contribution saturates. In particular for these scales, the purely contractile motion carries about $54 \%$ of the corresponding resolved-scale energy while the purely extensional stresses approximately $41\%$, \blue{as a result} the remaining part is due to the indefinite-type deformations. 

\section{Connecting resolved scale energy and enstrophy}
\purple{In this section,} as a complementary procedure, we are going to express the mean squared VG appearing on the RHS of \eqref{eq:eq_almost_final_ell} in terms of the resolved-scale enstrophy using \eqref{eq:first_betchov}. \blue{Hence, the equivalent expression to \eqref{eq:eq_final_ell_eig} yields:}
\begin{equation}
\lan E_K^\ell \ran =  \dfrac{1}{2} \int_{\ell^2}^{\infty} \hspace{-0.1cm}  \, \d \theta \, \lan \, || \overline{\bm{\omega}}^{\sqrt{\theta}} ||^2 \, \ran .
\label{eq:eq_final}
\end{equation}
which \blue{reveals} that, by measuring the resolved-scale enstrophy in real space at each scale, it is possible to retrieve the (resolved-scale) mean kinetic energy.

\purple{The prior equation} becomes \blue{definitely} more interesting in 2D turbulence where the enstrophy $\zeta = \frac{1}{2} \lan  || \overline{\bm{\omega}} ||^2 \ran$ is an inviscid conserved quantity; in this scenario the above equation connects two conserved quantities which is non-trivial. It is even less trivial that 
in 3D flows a conserved (inviscid) quantity can be expressed via the coarse-graining of a non-conserved one.
Eq.~\eqref{eq:eq_final} can be \blue{roughly thought of as} a sort of \emph{enstrophy-based quantization} of the kinetic energy in analogy with the Hamiltonian operator \blue{formulated via ladder operators in quantum field theory \cite{peskin2018}}. Connected to this interpretation, we underline that \blue{approximating the unresolved-scale energy} integral of eq.~\eqref{eq:eq_final} at a given scale $\ell$, meaning $\tau_{ii}^\ell \approx \frac{1}{2} \ell^2 ||\overline{\bm{\omega}}^\ell||^2$, resembles the expression of the \blue{rotational} energy of a ring of radius $\ell$ \blue{from rigid body mechanics.}

It is remarkable to notice that eq.~\eqref{eq:eq_final} and in consequence \eqref{eq:eq_final_ell_eig}, setting $\ell=0$ are more general. \purple{In order to discuss this}, \blue{we focus on the RHS of eq.~\eqref{eq:eq_final} 
disentangling it from its derivation. In this respect we consider the coarse-graining on the vorticity to be
}
implemented by a generic filter kernel $G^\ell$ instead of Gaussian kernel \purple{on} which the presented methodology is based. In this manner, the RHS of eq.~\eqref{eq:eq_final} downgrades to: 
\begin{align}
\dfrac{1}{2} \int_{\ell^2}^{\infty} \hspace{-0.1cm}  \, \d \theta \, \lan \, || \overline{\bm{\omega}}^{\sqrt{\theta}} ||^2 \, \ran  =&  \int_{\ell^2}^{\infty} \hspace{-0.1cm} \d \theta  \int_{\mathbb{R}^3}  \hspace{-0.22cm}  \,  \, \d \bm{k} \, k^2 \,\tilde{{u}}_i(\bm{k}) \, \tilde{{u}}^{*}_i \! (\bm{k}) \, \tilde{G}^{\sqrt{\theta}}(k)^2
\notag \\
=& \int_{\mathbb{R}^3}  \, \hspace{-0.15cm}  \,  \, \d \bm{k} \, \, \tilde{{u}}_i(\bm{k}) \, \tilde{{u}}_i^{*}\!(\bm{k})  \int_{\ell^2 k^2}^{\infty} \hspace{-0.1cm} \d s \, \tilde{G}(\sqrt{s})^2
\label{eq:michael_lukas}   
\end{align}
where \blue{$(\cdot)^*$ denotes the complex conjugation}. For simplicity, we assume that the velocity field allows the continuous Fourier representation (see \cite{frisch1995} for details). Back to the RHS of \eqref{eq:michael_lukas}, the kernel dependency on the filtering-scale has been absorbed into its argument i.e. $\tilde{G}^{\sqrt{\theta}}(k) = \tilde{G}(k \sqrt{\theta})$. This effective rescaling holds for the standard filter functions, like those listed in table 13.2 of \citep{pope2001} but in principle  this would remove from our discussion \emph{exotic} filter kernel types for which the \blue{preceding rescaling} does not hold. \purple{Like already mentioned above, the inversion of the integrals in the derivation of \eqref{eq:michael_lukas} is a consequence of the assumed convergence of the fields involved.}
 
\blue{At this stage, we observe \citep{lukasidea,michaelchat} that by setting} $\ell=0$ in \eqref{eq:michael_lukas}, the two integrals can be factorised. In this way the RHS of \eqref{eq:michael_lukas} collapses \blue{in}to the product $C_G\, \lan E_K\ran$ where $C_G$ is a real number depending on the filter choice \blue{and the mean energy is given by the integration of the velocity Fourier modes defined in \eqref{eq:fourier}}. In such a way, we have generalised eq.~\eqref{eq:eq_final} to a generic class of filters without introducing the SGS stresses decomposition of \eqref{eq:perrys_meth} and without requiring the associated positive definiteness of its trace. Note that a similar factorization can be derived for the $nth$-derivative of the velocity field \cite{michaelchat}.

However, \blue{depending on the filter kernel,} the generalisation originating from the preceding observation does not necessarily hold when \eqref{eq:michael_lukas} is evaluated at $\ell \neq 0$\blue{; this condition prevents the calculation of the related kinetic energy spectrum in eq.~\eqref{eq:en_el_to_spec}}. In order to show this, we consider the \emph{sharp filter} whose Fourier transform is a Heaviside function $G^\ell(k) = H(\pi - k \,\ell)$. For this kernel, eq.~\eqref{eq:michael_lukas} becomes $\int_0^{\pi/\ell} \d k\, E(k) (\pi^2 -\ell^2 k^2)$ which does not equal \blue{the corresponding} mean resolved-scale energy $\lan E^\ell_K \ran = \int_0^{\pi/\ell} \d k \, E(k) $.

\section{Addressing the non-stationarity}


Having obtained an exact expansion of the mean kinetic energy in \eqref{eq:eq_final} allows for the dynamical characterization of non-stationary flows $\partial_t \langle E_K (t) \rangle \gtrless 0$, respectively describing the flow energy increase and the turbulence decay where the time dependence of the energy is purposefully reinstated. For this reason, applying a time derivative to eq.~\eqref{eq:eq_almost_final} one can obtain an expression involving \blue{filtered} \blue{VG} quantities. For simplicity we focus on the homogeneous turbulence case, where the calculation of the time derivative in \eqref{eq:eq_almost_final} reads:
\begin{align}
&\partial_t \langle E_K(t) \rangle  = \frac{1}{2}\int_0^{\infty} \d \theta \, \partial_t \, \langle  \overline{u}_{i,j}^{\sqrt{\theta}} \overline{u}_{i,j}^{\sqrt{\theta}}   \rangle \label{eq:non_stat_erg_int}  \\
&\hspace{-0.18cm}= \hspace{-0.1cm} \int_0^{\infty} \hspace{-0.2cm} \d \theta \,  \Big( -\dfrac{4}{3} \lan  \overline{S}_{ij}^{\sqrt{\theta}} \overline{S}_{jk}^{\sqrt{\theta}} \overline{S}_{ki}^{\sqrt{\theta}}  \ran  - \nu \lan \overline{u}^{\sqrt{\theta}}_{i,jk} \, \overline{u}^{\sqrt{\theta}}_{i,jk} \ran - \lan \overline{S}^{\sqrt{\theta}}_{ij,\,kk} \tau_{ij}^{\sqrt{\theta}} \ran \notag \\
&\hspace{-0.1cm} + \lan F^{\sqrt{\theta}} \ran \hspace{-0.01cm}  \bigg)
\label{eq:non_stat_erg}
\end{align}  
where $F^{\sqrt{\theta}} = \overline{u}^{\sqrt{\theta}}_{i,j} \overline{f}^{\sqrt{\theta}}_{i,j}$ depends on the external forcing and it is set to zero in the case of turbulence decay. As concerns the other terms, the one containing the contraction of three strain-rate tensors is linked to a contribution of \emph{energy transfer} across the scales \cite{johnson2020}. Numerical evidence \cite{johnson2020}\cite{johnson2021} shows that it is positive $\forall\, \ell$ at the stationary state, then \blue{it is expected to} maintain the same sign in the present non-stationary case as well. \blue{As for the third term}, it is formed by the contraction between the strain-rate tensor Laplacian and the SGS stresses; being a priori sign indefinite, it may play a relevant role in the above energy evolution equation. Such a term also shows an intrinsic \emph{multi-scale} nature from the presence of the SGS stress tensor which, by virtue of \eqref{eq:perrys_meth}, depends on the scales smaller than the integration scale $\sqrt{\theta}$. Finally, it is clear that the term proportional to the viscosity is negative. Applying \eqref{eq:non_stat_erg} to data from DNS will assess the importance of such terms in unsteady configurations. 

Independently from the quantification of the terms, in case of stationarity i.e. $\partial_t \lan E_K \ran = 0$, the above expression indicates that the stationarity can, in principle, be reached in two ways: a scale-by-scale stationarity where the contribution of each integrated scale is identically zero or the \emph{scale-time ergodicity} \cite{moritzsidea} where the RHS of \eqref{eq:non_stat_erg_int} is zero because of a compensation of the non-stationarity of the corresponding scales contributions. 

It is worth observing that in 2D turbulence eq.~\eqref{eq:non_stat_erg} simplifies dramatically.  Firstly the contraction of the three (resolved scale) strain-rate tensors is identically zero pointwise (see e.g. Appendix C.1 of \cite{damianinoMHD2024}). Secondly, in the third term, the decomposition of the SGS stresses in \eqref{eq:perrys_meth} comprises of only one term \footnote{Given a symmetric tensor $A_{ij}$, in 2D one can show that $A_{ij} \tau_{ij}^{\ell} = A_{ij}  \int_{0}^{\ell^2} \hspace{-0.1cm}  \, \d q \, \overline{ \, \overline{S}^{\sqrt{q}}_{ij} \, \overline{\Omega}^{\sqrt{q}}_{ij} }^{\sqrt{\ell^2 - q}}$}\cite{johnson2021}, as opposed to the general 3D case. \purple{It follows that eq.~\eqref{eq:non_stat_erg} can help to identify the governing physical mechanisms that lead the the turbulence decay or the increase of energy}.

\section{An exact identity for the kinetic energy spectrum}

From the knowledge of the resolved-scale kinetic energy it is possible to obtain the kinetic energy spectrum \blue{$E(k)$} through the following formula \footnote{{By expressing the total resolved-scale kinetic energy as the sum of the energy spectrum over the corresponding wavenumber interval $\int_0^{k} \d k' \,E(k') = \lan E^{\pi/k} \ran $ (a special case of eq. (6.14) of \cite{pope2001}) one can derive $E(k) = \dfrac{\p \lan E^{\pi/k} \ran}{\p k}$, which is subsequently written and manipulated in terms of the filter scale $\ell=\pi/k$.}}:
\begin{equation}
E(k) = -\dfrac{\pi}{k^2} \dfrac{\p}{\p \ell} \lan E^\ell_K \rangle  
\label{eq:en_el_to_spec}
\end{equation}
where \blue{the time dependence has been omitted}. In order to get an expression in terms of velocity gradient, we can plug eq.~\eqref{eq:eq_final} into the previous equation obtaining:
\begin{equation}
E(k) = \pi^2 k^{-3} \, \lan || \bm{\overline{\omega}}^{\pi/k} ||^2  \ran 
\label{eq:simil_kolmogorov}
\end{equation}
which is an exact expression for the kinetic energy spectrum where the wavenumber dependency is both kept by the \blue{wavenumber prefactor} and by the filter scale that the vorticity is coarse-grained at. 

Similar to the discussion regarding eq.~\eqref{eq:eq_almost_final}, when eq.~\eqref{eq:simil_kolmogorov} is evaluated at \emph{high} wavenumbers (i.e. at small-scales) the \emph{large} values of the squared vorticity norm are penalised by the ${k^{-3}}$ prefactor. Conversely, at \emph{low} wavenumbers (large-scales) the prefactor exalts the resulting \emph{smaller} values of the squared vorticity norm in the overall product.

\purple{The above equation must be compared with the \emph{usual} definition of the kinetic energy spectrum:}
\begin{equation}
E^U(k) =  \dfrac{1}{2} \int_{\mathbb{R}^3} \d \bm{k} \,\, \tilde{u}_i(\bm{k})\, \tilde{u}^{*}_i(\bm{k}) \, \delta(||\bm{k}|| - k)
\label{eq:spec} 
\end{equation}
where $\delta(\cdot)$ is a delta function. \blue{It follows that eq.~\eqref{eq:simil_kolmogorov} qualifies as an alternative \emph{representation} of the kinetic energy spectrum} \footnote{\blue{By expressing the filtered enstrophy in Fourier space, one can notice that the} rationale of eq.~\eqref{eq:simil_kolmogorov} consists in replacing the delta function in \eqref{eq:spec} by a unit-measure Gaussian.} \blue{which,} unlike eq.~\eqref{eq:spec}, \blue{can} be determined directly by real space quantities like the resolved-scale enstrophy. \blue{Thus, we can consider the dataset used so far to test eq.~\eqref{eq:simil_kolmogorov} by comparing it with the common definition of energy spectrum from eq.~\eqref{eq:spec}.} Moreover, in turbulence theory, when the study of energy spectra is of interest, it is customary to discuss the corresponding Kolmogorov's scaling \citep{kolmogorov1941a,kolmogorov1941b,frisch1995} which is $E^{KOL}(k) = C \, \lan \varepsilon \ran^{2/3} \, k^{-5/3} $ where $C$ is the Kolmogorov constant, according to the employed dataset its value is nearly $1.6$, see \cite{damianinoHDdata} \blue{for details}. 

\begin{figure}
	\begin{center}
         \includegraphics[width=1.1\columnwidth]{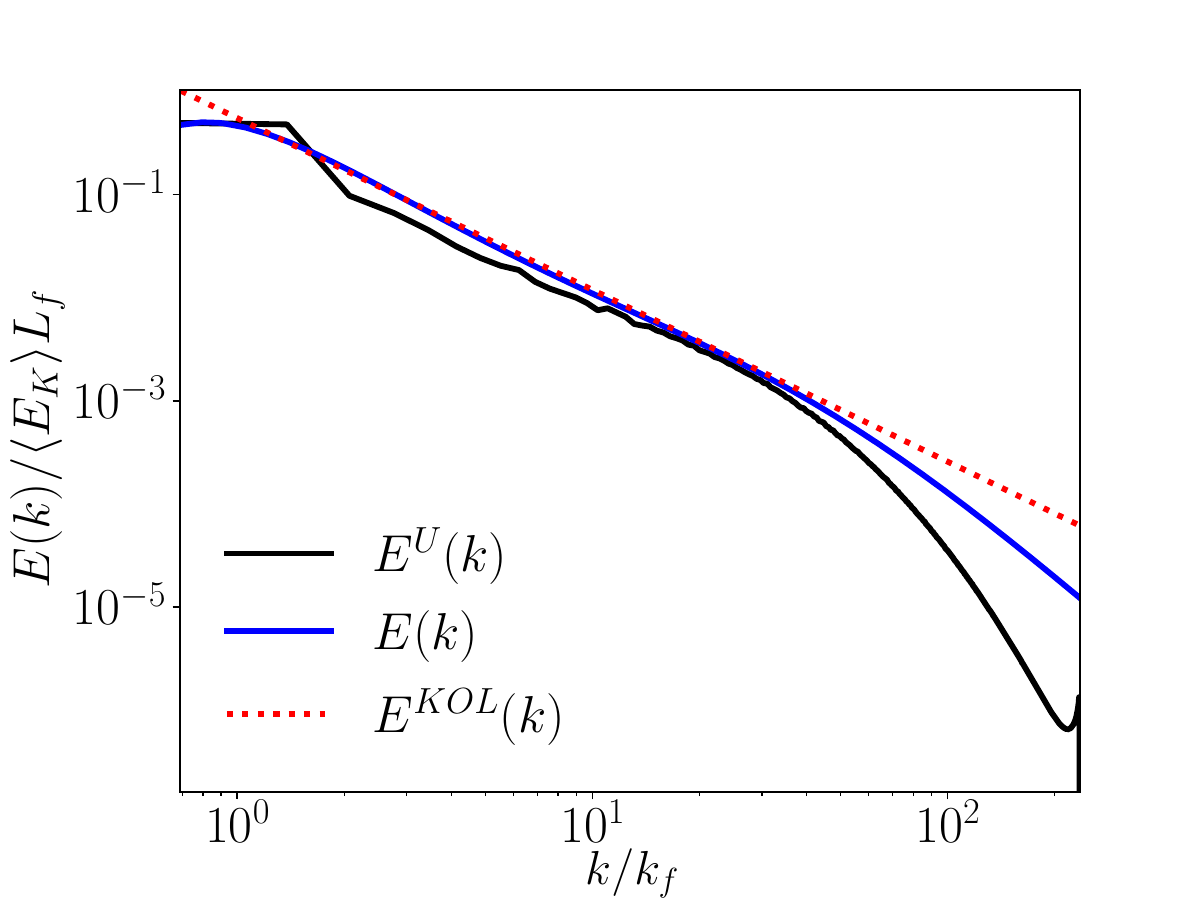} 
        \end{center}
	 \caption{Omni-directional kinetic energy spectrum \blue{based on} one instantaneous configuration and calculated respectively via the two different \blue{representations of eqs.~\eqref{eq:simil_kolmogorov} and \eqref{eq:spec}}. The dotted line refers to the Kolmogorov power-law spectral scaling. \blue{All the curves are expressed as functions of the non-dimensional quantity $k/k_f=L_f/\ell$ and normalised by $\langle E_K \rangle \, L_f$.} }
	 \label{fig:spec_comp}
\end{figure}

Fig.~\ref{fig:spec_comp} shows the comparison between eqs.~\eqref{eq:spec}, \eqref{eq:simil_kolmogorov} and the Kolmogorov scaling in the evaluation of the kinetic energy spectrum.
The first feature we notice is the good agreement \blue{of $E(k)$} with Kolmogorov scaling \footnote{Note that by \cite{eyink1995}, $|| \overline{\bm{\omega}}^{\ell} || \sim \ell^{-2/3}$ hence eq.~\eqref{eq:simil_kolmogorov} scales as $ \ell^{5/3} \sim k^{-5/3}$ like the Kolmogorov formula.} in the wavenumber band $ 1.5 \lesssim  k/k_f \lesssim 30.0\,$. As concerns the comparison between $E(k)$ and $E^U(k)$, we appreciate that the former reproduces the latter in the forcing range i.e. $k/k_f \approx 1$. It is intuitive to recognize both the smoother profile of $E(k)$ and the discrepancy for large $k$ with $E^U(k)$ as an \emph{inheritance} of the Gaussian filter, according \blue{to} which filtering at wavenumber $k$ still retains features from $< k$.

For those scales where $E^{KOL}(k)=E(k)$, we can equate the two corresponding expressions tentatively writing an expression for the Kolmogorov constant:
\begin{equation}
C = \pi^2 \, \dfrac{\lan || \bm{\overline{\omega}}^{\pi/k} ||^2  \ran}{\lan \varepsilon \ran^{2/3} \, k^{4/3}  }
\label{eq:kolm_const}
\end{equation} 
that requires further investigations.
However a rough estimate would be evaluating \eqref{eq:kolm_const} at $k/k_f \gg 1$, for instance at the so-called Kolmogorov microscale $\eta = (\nu^3/\varepsilon)^{1/4}$ s.t. $\eta \ll L_f $ \footnote{The \blue{approximation} of this calculation \blue{is based} on equating the Kolmogorov scaling $E^{KOL}(k)$ with $E(k)$ of eq.~\eqref{eq:simil_kolmogorov} at Kolmogorov micro-scale $\eta=(\nu^3/\lan \varepsilon \rangle)^{1/4}$ for which $k_\eta/k_f= 344$. Even though this wavenumber goes beyond the resolution of the analysed dataset, fig.~\ref{fig:spec_comp} potentially indicates that the two curves do not intersect. From an analytical point of view, \emph{large} values of $k$ would make \eqref{eq:kolm_const} independent of both the viscosity $\nu$ the (resolved-scale) enstrophy where the latter is due to the legitimacy of the approximation $ ||\bm{\overline{\omega}}^{\pi/k} || \approx || \bm{{\omega}}|| $. }. This approximation would provide a $C=\pi^{2/3}\simeq 2.1$ against a $C\approx 1.6$ measured from the compensated energy spectrum \cite{damianinoHDdata}. See \cite{donzis2010} and references therein for a complete discussion on \blue{characterisation} of the Kolmogorov constant.

\section{Conclusions and future directions}

In conclusion, from a more general identity relating the total kinetic energy and the coarse-grained velocity gradient norm based on the Gaussian filter kernel, two (sub)identities are derived for homogeneous turbulence. One expresses the mean energy balance in terms \blue{of the strain-rate intensity}. Outside the range of scale dominated by the forcing, the purely contractile and extensional deformations provide \blue{respectively} nearly $55\%$ and $40\%$ of the resolved-scale energy. The remainder is linked to the intermediate eigenvalues which can be either related to contractions or extensions. Using kinematic constraints, one can obtain a cognate identity 
showing that \blue{in turn} the sum of \blue{the} enstrophy across the scales determines the total kinetic energy. Finally, one can obtain an exact identity for the kinetic energy spectrum in terms of in real space interpretable observables providing a deeper fundamental understanding of the energy spectrum and a guidance for the analytical determination of the Kolmogorov constant. \blue{As a consequence of the properties of SGS stresses expansion from \eqref{eq:perrys_meth}, this present methodology can be applied to any process characterised by an advective-type non-linearity. On top of that, beyond the strain-vorticity splitting hereby considered, we can also perform different VG decompositions, e.g. \cite{kolar2007} \cite{das2020}, to further explore the role of the VG norm in the energy budget across the scales}. The analysis of the dynamical equation describing unsteady flows can be deeply expanded to unveil the governing physical mechanisms that lead the turbulence decay or the increase of energy.


\section{Acknowledgments and fundings}

The author thanks Prof. Luca Biferale and Dr. Massimo Cencini for the financial support provided by both European Research Council (ERC) under the European Union's Horizon 2020 research and innovation programme (grant agreement No 882340) and FieldTurb experiment of the Istituto Nazionale di Fisica Nucleare (INFN). This work used the ARCHER2 UK National Supercomputing Service ({\tt www.archer2.ac.uk}) with resources provided by the UK Turbulence Consortium (EPSRC grants EP/R029326/1). The author thanks Dr. Moritz Linkmann for the interesting discussions, encouragement and the support. The author wants to thank Prof. Michael Wilczek and Lukas Bentkamp for their valuable observations and Dr. Martin Lellep for  editing suggestions.

\bibliography{apssamp}

 \providecommand{\SortNoop}[1]{} 
  \providecommand{\sortnoop}[1]{} 
\begin{thebibliography}{37}%
\makeatletter
\providecommand \@ifxundefined [1]{%
 \@ifx{#1\undefined}
}%
\providecommand \@ifnum [1]{%
 \ifnum #1\expandafter \@firstoftwo
 \else \expandafter \@secondoftwo
 \fi
}%
\providecommand \@ifx [1]{%
 \ifx #1\expandafter \@firstoftwo
 \else \expandafter \@secondoftwo
 \fi
}%
\providecommand \natexlab [1]{#1}%
\providecommand \enquote  [1]{``#1''}%
\providecommand \bibnamefont  [1]{#1}%
\providecommand \bibfnamefont [1]{#1}%
\providecommand \citenamefont [1]{#1}%
\providecommand \href@noop [0]{\@secondoftwo}%
\providecommand \href [0]{\begingroup \@sanitize@url \@href}%
\providecommand \@href[1]{\@@startlink{#1}\@@href}%
\providecommand \@@href[1]{\endgroup#1\@@endlink}%
\providecommand \@sanitize@url [0]{\catcode `\\12\catcode `\$12\catcode
  `\&12\catcode `\#12\catcode `\^12\catcode `\_12\catcode `\%12\relax}%
\providecommand \@@startlink[1]{}%
\providecommand \@@endlink[0]{}%
\providecommand \url  [0]{\begingroup\@sanitize@url \@url }%
\providecommand \@url [1]{\endgroup\@href {#1}{\urlprefix }}%
\providecommand \urlprefix  [0]{URL }%
\providecommand \Eprint [0]{\href }%
\providecommand \doibase [0]{https://doi.org/}%
\providecommand \selectlanguage [0]{\@gobble}%
\providecommand \bibinfo  [0]{\@secondoftwo}%
\providecommand \bibfield  [0]{\@secondoftwo}%
\providecommand \translation [1]{[#1]}%
\providecommand \BibitemOpen [0]{}%
\providecommand \bibitemStop [0]{}%
\providecommand \bibitemNoStop [0]{.\EOS\space}%
\providecommand \EOS [0]{\spacefactor3000\relax}%
\providecommand \BibitemShut  [1]{\csname bibitem#1\endcsname}%
\let\auto@bib@innerbib\@empty
\bibitem [{\citenamefont {Buaria}\ \emph {et~al.}(2019)\citenamefont {Buaria},
  \citenamefont {Pumir}, \citenamefont {Bodenschatz},\ and\ \citenamefont
  {Yeung}}]{buaria2019}%
  \BibitemOpen
  \bibfield  {author} {\bibinfo {author} {\bibfnamefont {D.}~\bibnamefont
  {Buaria}}, \bibinfo {author} {\bibfnamefont {A.}~\bibnamefont {Pumir}},
  \bibinfo {author} {\bibfnamefont {E.}~\bibnamefont {Bodenschatz}},\ and\
  \bibinfo {author} {\bibfnamefont {P.-K.}\ \bibnamefont {Yeung}},\ }\bibfield
  {title} {\bibinfo {title} {Extreme velocity gradients in turbulent flows},\
  }\href@noop {} {\bibfield  {journal} {\bibinfo  {journal} {New Journal of
  Physics}\ }\textbf {\bibinfo {volume} {21}},\ \bibinfo {pages} {043004}
  (\bibinfo {year} {2019})}\BibitemShut {NoStop}%
\bibitem [{\citenamefont {Thorpe}(2005)}]{thorpe2005}%
  \BibitemOpen
  \bibfield  {author} {\bibinfo {author} {\bibfnamefont {S.~A.}\ \bibnamefont
  {Thorpe}},\ }\href@noop {} {\emph {\bibinfo {title} {The turbulent ocean}}}\
  (\bibinfo  {publisher} {Cambridge university press},\ \bibinfo {year}
  {2005})\BibitemShut {NoStop}%
\bibitem [{\citenamefont {Pedlosky}(2013)}]{pedlosky2013}%
  \BibitemOpen
  \bibfield  {author} {\bibinfo {author} {\bibfnamefont {J.}~\bibnamefont
  {Pedlosky}},\ }\href@noop {} {\emph {\bibinfo {title} {Geophysical fluid
  dynamics}}}\ (\bibinfo  {publisher} {Springer Science \& Business Media},\
  \bibinfo {year} {2013})\BibitemShut {NoStop}%
\bibitem [{\citenamefont {Cho}\ \emph {et~al.}(2003)\citenamefont {Cho},
  \citenamefont {Lazarian},\ and\ \citenamefont {Vishniac}}]{cho2003}%
  \BibitemOpen
  \bibfield  {author} {\bibinfo {author} {\bibfnamefont {J.}~\bibnamefont
  {Cho}}, \bibinfo {author} {\bibfnamefont {A.}~\bibnamefont {Lazarian}},\ and\
  \bibinfo {author} {\bibfnamefont {E.~T.}\ \bibnamefont {Vishniac}},\
  }\href@noop {} {\emph {\bibinfo {title} {MHD turbulence: scaling laws and
  astrophysical implications}}}\ (\bibinfo  {publisher} {Springer},\ \bibinfo
  {year} {2003})\BibitemShut {NoStop}%
\bibitem [{\citenamefont {Goldstein}\ \emph {et~al.}(1995)\citenamefont
  {Goldstein}, \citenamefont {Roberts},\ and\ \citenamefont
  {Matthaeus}}]{goldstein1995}%
  \BibitemOpen
  \bibfield  {author} {\bibinfo {author} {\bibfnamefont {M.~L.}\ \bibnamefont
  {Goldstein}}, \bibinfo {author} {\bibfnamefont {D.~A.}\ \bibnamefont
  {Roberts}},\ and\ \bibinfo {author} {\bibfnamefont {W.}~\bibnamefont
  {Matthaeus}},\ }\bibfield  {title} {\bibinfo {title} {Magnetohydrodynamic
  turbulence in the solar wind},\ }\href@noop {} {\bibfield  {journal}
  {\bibinfo  {journal} {Annual review of astronomy and astrophysics}\ }\textbf
  {\bibinfo {volume} {33}},\ \bibinfo {pages} {283} (\bibinfo {year}
  {1995})}\BibitemShut {NoStop}%
\bibitem [{\citenamefont {Goldreich}\ and\ \citenamefont
  {Sridhar}(1997)}]{goldreich1997}%
  \BibitemOpen
  \bibfield  {author} {\bibinfo {author} {\bibfnamefont {P.}~\bibnamefont
  {Goldreich}}\ and\ \bibinfo {author} {\bibfnamefont {S.}~\bibnamefont
  {Sridhar}},\ }\bibfield  {title} {\bibinfo {title} {Magnetohydrodynamic
  turbulence revisited},\ }\href@noop {} {\bibfield  {journal} {\bibinfo
  {journal} {The Astrophysical Journal}\ }\textbf {\bibinfo {volume} {485}},\
  \bibinfo {pages} {680} (\bibinfo {year} {1997})}\BibitemShut {NoStop}%
\bibitem [{\citenamefont {Germano}(1992)}]{germano1992}%
  \BibitemOpen
  \bibfield  {author} {\bibinfo {author} {\bibfnamefont {M.}~\bibnamefont
  {Germano}},\ }\bibfield  {title} {\bibinfo {title} {Turbulence: the filtering
  approach},\ }\href@noop {} {\bibfield  {journal} {\bibinfo  {journal}
  {Journal of Fluid Mechanics}\ }\textbf {\bibinfo {volume} {238}},\ \bibinfo
  {pages} {325} (\bibinfo {year} {1992})}\BibitemShut {NoStop}%
\bibitem [{\citenamefont {Pope}(2000)}]{pope2001}%
  \BibitemOpen
  \bibfield  {author} {\bibinfo {author} {\bibfnamefont {S.~B.}\ \bibnamefont
  {Pope}},\ }\href@noop {} {\emph {\bibinfo {title} {Turbulent flows}}}\
  (\bibinfo  {publisher} {Cambridge University Press},\ \bibinfo {year}
  {2000})\BibitemShut {NoStop}%
\bibitem [{\citenamefont {Vreman}\ \emph {et~al.}(1994)\citenamefont {Vreman},
  \citenamefont {Geurts},\ and\ \citenamefont {Kuerten}}]{vreman1994}%
  \BibitemOpen
  \bibfield  {author} {\bibinfo {author} {\bibfnamefont {B.}~\bibnamefont
  {Vreman}}, \bibinfo {author} {\bibfnamefont {B.}~\bibnamefont {Geurts}},\
  and\ \bibinfo {author} {\bibfnamefont {H.}~\bibnamefont {Kuerten}},\
  }\bibfield  {title} {\bibinfo {title} {Realizability conditions for the
  turbulent stress tensor in large-eddy simulation},\ }\href@noop {} {\bibfield
   {journal} {\bibinfo  {journal} {Journal of Fluid Mechanics}\ }\textbf
  {\bibinfo {volume} {278}},\ \bibinfo {pages} {351} (\bibinfo {year}
  {1994})}\BibitemShut {NoStop}%
\bibitem [{Note1()}]{Note1}%
  \BibitemOpen
  \bibinfo {note} {Note that, for an unbounded domain, the Fourier
  representation can be bypassed by requiring that $g(\protect \bm {x})$ goes
  to zero as $||\protect \bm {x}|| \to \infty $. The bypass is formal in
  periodic boundaries which in turn would imply the expansion in Fourier
  series. The same considerations applies to the derivation of eq.~\protect
  \eqref {eq:perrys_meth} in case one wants to avoid the Fourier
  expansion.}\BibitemShut {Stop}%
\bibitem [{\citenamefont {Aluie}(2017)}]{aluie2017}%
  \BibitemOpen
  \bibfield  {author} {\bibinfo {author} {\bibfnamefont {H.}~\bibnamefont
  {Aluie}},\ }\bibfield  {title} {\bibinfo {title} {Coarse-grained
  incompressible magnetohydrodynamics: analyzing the turbulent cascades},\
  }\href@noop {} {\bibfield  {journal} {\bibinfo  {journal} {New Journal of
  Physics}\ }\textbf {\bibinfo {volume} {19}},\ \bibinfo {pages} {025008}
  (\bibinfo {year} {2017})}\BibitemShut {NoStop}%
\bibitem [{\citenamefont {Zinn-Justin}(2010)}]{zinn2010}%
  \BibitemOpen
  \bibfield  {author} {\bibinfo {author} {\bibfnamefont {J.}~\bibnamefont
  {Zinn-Justin}},\ }\href@noop {} {\emph {\bibinfo {title} {Path integrals in
  quantum mechanics}}}\ (\bibinfo  {publisher} {OUP Oxford},\ \bibinfo {year}
  {2010})\BibitemShut {NoStop}%
\bibitem [{\citenamefont {Johnson}(2020)}]{johnson2020}%
  \BibitemOpen
  \bibfield  {author} {\bibinfo {author} {\bibfnamefont {P.~L.}\ \bibnamefont
  {Johnson}},\ }\bibfield  {title} {\bibinfo {title} {Energy transfer from
  large to small scales in turbulence by multiscale nonlinear strain and
  vorticity interactions},\ }\href
  {https://doi.org/10.1103/PhysRevLett.124.104501} {\bibfield  {journal}
  {\bibinfo  {journal} {Phys.\ Rev.\ Lett.}\ }\textbf {\bibinfo {volume}
  {124}},\ \bibinfo {pages} {104501} (\bibinfo {year} {2020})}\BibitemShut
  {NoStop}%
\bibitem [{\citenamefont {Johnson}(2021)}]{johnson2021}%
  \BibitemOpen
  \bibfield  {author} {\bibinfo {author} {\bibfnamefont {P.~L.}\ \bibnamefont
  {Johnson}},\ }\bibfield  {title} {\bibinfo {title} {On the role of vorticity
  stretching and strain self-amplification in the turbulence energy cascade},\
  }\href@noop {} {\bibfield  {journal} {\bibinfo  {journal} {Journal of Fluid
  Mechanics}\ }\textbf {\bibinfo {volume} {922}},\ \bibinfo {pages} {A3}
  (\bibinfo {year} {2021})}\BibitemShut {NoStop}%
\bibitem [{Note2()}]{Note2}%
  \BibitemOpen
  \bibinfo {note} {Note that the methodology of \cite {johnson2020} requires
  homogeneity or periodic boundaries in order to set to zero the boundary terms
  and not for the derivation of the decomposition \protect \eqref
  {eq:perrys_meth} itself.}\BibitemShut {Stop}%
\bibitem [{Note3()}]{Note3}%
  \BibitemOpen
  \bibinfo {note} {The length contraction in general relativity would already
  pose conceptual difficulties to the present filtering approach.}\BibitemShut
  {Stop}%
\bibitem [{Note4()}]{Note4}%
  \BibitemOpen
  \bibinfo {note} {However, in the applications, the interest \textcolor
  {black}{clearly} focuses on systems characterised by a maximum allowed scale
  for which the integration upper bound of \protect \eqref {eq:perrys_meth}
  does not diverge ensuring a rigorous exchange of integrals to derive \protect
  \eqref {eq:eq_interm}.}\BibitemShut {Stop}%
\bibitem [{\citenamefont {Biferale}\ \emph {et~al.}(2024)\citenamefont
  {Biferale}, \citenamefont {Bonaccorso}, \citenamefont {Linkmann},\ and\
  \citenamefont {Capocci}}]{damianinoHDdata}%
  \BibitemOpen
  \bibfield  {author} {\bibinfo {author} {\bibfnamefont {L.}~\bibnamefont
  {Biferale}}, \bibinfo {author} {\bibfnamefont {F.}~\bibnamefont
  {Bonaccorso}}, \bibinfo {author} {\bibfnamefont {M.}~\bibnamefont
  {Linkmann}},\ and\ \bibinfo {author} {\bibfnamefont {D.}~\bibnamefont
  {Capocci}},\ }\bibfield  {title} {\bibinfo {title} {Turb-hel: an open-access
  database of helically forced homogeneous and isotropic turbulence},\
  }\href@noop {} {\bibfield  {journal} {\bibinfo  {journal} {arXiv preprint
  arXiv:2404.07653}\ } (\bibinfo {year} {2024})}\BibitemShut {NoStop}%
\bibitem [{\citenamefont {Tennekes}\ and\ \citenamefont
  {Lumley}(1972)}]{tennekes1972}%
  \BibitemOpen
  \bibfield  {author} {\bibinfo {author} {\bibfnamefont {H.}~\bibnamefont
  {Tennekes}}\ and\ \bibinfo {author} {\bibfnamefont {J.~L.}\ \bibnamefont
  {Lumley}},\ }\href@noop {} {\emph {\bibinfo {title} {A first course in
  turbulence}}}\ (\bibinfo  {publisher} {MIT press},\ \bibinfo {year}
  {1972})\BibitemShut {NoStop}%
\bibitem [{\citenamefont {Betchov}(1956)}]{betchov1956}%
  \BibitemOpen
  \bibfield  {author} {\bibinfo {author} {\bibfnamefont {R.}~\bibnamefont
  {Betchov}},\ }\bibfield  {title} {\bibinfo {title} {An inequality concerning
  the production of vorticity in isotropic turbulence},\ }\href@noop {}
  {\bibfield  {journal} {\bibinfo  {journal} {Journal of Fluid Mechanics}\
  }\textbf {\bibinfo {volume} {1}},\ \bibinfo {pages} {497} (\bibinfo {year}
  {1956})}\BibitemShut {NoStop}%
\bibitem [{\citenamefont {Peskin}(2018)}]{peskin2018}%
  \BibitemOpen
  \bibfield  {author} {\bibinfo {author} {\bibfnamefont {M.~E.}\ \bibnamefont
  {Peskin}},\ }\href@noop {} {\emph {\bibinfo {title} {An introduction to
  quantum field theory}}}\ (\bibinfo  {publisher} {CRC press},\ \bibinfo {year}
  {2018})\BibitemShut {NoStop}%
\bibitem [{\citenamefont {Frisch}(1995)}]{frisch1995}%
  \BibitemOpen
  \bibfield  {author} {\bibinfo {author} {\bibfnamefont {U.}~\bibnamefont
  {Frisch}},\ }\href@noop {} {\emph {\bibinfo {title} {Turbulence: the legacy
  of AN Kolmogorov}}}\ (\bibinfo  {publisher} {Cambridge University Press},\
  \bibinfo {year} {1995})\BibitemShut {NoStop}%
\bibitem [{\citenamefont {Bentkamp}()}]{lukasidea}%
  \BibitemOpen
  \bibfield  {author} {\bibinfo {author} {\bibfnamefont {L.}~\bibnamefont
  {Bentkamp}},\ }\href@noop {} {\bibinfo {title} {Private
  communication}}\BibitemShut {NoStop}%
\bibitem [{\citenamefont {Wilczek}()}]{michaelchat}%
  \BibitemOpen
  \bibfield  {author} {\bibinfo {author} {\bibfnamefont {M.}~\bibnamefont
  {Wilczek}},\ }\href@noop {} {\bibinfo {title} {Private
  communication}}\BibitemShut {NoStop}%
\bibitem [{\citenamefont {Linkmann}()}]{moritzsidea}%
  \BibitemOpen
  \bibfield  {author} {\bibinfo {author} {\bibfnamefont {M.}~\bibnamefont
  {Linkmann}},\ }\href@noop {} {\bibinfo {title} {Private
  communication}}\BibitemShut {NoStop}%
\bibitem [{\citenamefont {Capocci}\ \emph {et~al.}(2025)\citenamefont
  {Capocci}, \citenamefont {Johnson}, \citenamefont {Oughton}, \citenamefont
  {Biferale},\ and\ \citenamefont {Linkmann}}]{damianinoMHD2024}%
  \BibitemOpen
  \bibfield  {author} {\bibinfo {author} {\bibfnamefont {D.}~\bibnamefont
  {Capocci}}, \bibinfo {author} {\bibfnamefont {P.~L.}\ \bibnamefont
  {Johnson}}, \bibinfo {author} {\bibfnamefont {S.}~\bibnamefont {Oughton}},
  \bibinfo {author} {\bibfnamefont {L.}~\bibnamefont {Biferale}},\ and\
  \bibinfo {author} {\bibfnamefont {M.}~\bibnamefont {Linkmann}},\ }\bibfield
  {title} {\bibinfo {title} {Energy flux decomposition in magnetohydrodynamic
  turbulence},\ }\href {https://doi.org/10.1017/S0022377824000898} {\bibfield
  {journal} {\bibinfo  {journal} {Journal of Plasma Physics}\ }\textbf
  {\bibinfo {volume} {91}},\ \bibinfo {pages} {E11} (\bibinfo {year}
  {2025})}\BibitemShut {NoStop}%
\bibitem [{Note5()}]{Note5}%
  \BibitemOpen
  \bibinfo {note} {Given a symmetric tensor $A_{ij}$, in 2D one can show that
  $A_{ij} \tau _{ij}^{\ell } = A_{ij} \DOTSI \intop \ilimits@ _{0}^{\ell ^2}
  \protect \hspace {-0.1cm} \protect \, \protect \mathrm {d}q \protect \,
  \protect \overline { \protect \, \protect \overline {S}^{\protect \sqrt
  {q}}_{ij} \protect \, \protect \overline {\Omega }^{\protect \sqrt {q}}_{ij}
  }^{\protect \sqrt {\ell ^2 - q}}$}\BibitemShut {NoStop}%
\bibitem [{Note6()}]{Note6}%
  \BibitemOpen
  \bibinfo {note} {{By expressing the total resolved-scale kinetic energy as
  the sum of the energy spectrum over the corresponding wavenumber interval
  $\DOTSI \intop \ilimits@ _0^{k} \protect \mathrm {d}k' \protect \,E(k') =
  \langle E^{\pi /k} \rangle $ (a special case of eq. (6.14) of \cite
  {pope2001}) one can derive $E(k) = \protect \dfrac {\partial \langle E^{\pi
  /k} \rangle }{\partial k}$, which is subsequently written and manipulated in
  terms of the filter scale $\ell =\pi /k$.}}\BibitemShut {Stop}%
\bibitem [{Note7()}]{Note7}%
  \BibitemOpen
  \bibinfo {note} {\textcolor {black}{By expressing the filtered enstrophy in
  Fourier space, one can notice that the} rationale of eq.~\protect \eqref
  {eq:simil_kolmogorov} consists in replacing the delta function in \protect
  \eqref {eq:spec} by a unit-measure Gaussian.}\BibitemShut {Stop}%
\bibitem [{\citenamefont {{Kolmogorov}}(1941)}]{kolmogorov1941a}%
  \BibitemOpen
  \bibfield  {author} {\bibinfo {author} {\bibfnamefont {A.~N.}\ \bibnamefont
  {{Kolmogorov}}},\ }\bibfield  {title} {\bibinfo {title} {{Dissipation of
  energy in locally isotropic turbulence}},\ }\href@noop {} {\bibfield
  {journal} {\bibinfo  {journal} {Akademiia Nauk SSSR Doklady}\ }\textbf
  {\bibinfo {volume} {32}},\ \bibinfo {pages} {16} (\bibinfo {year}
  {1941})}\BibitemShut {NoStop}%
\bibitem [{\citenamefont {{Kolmogorov}}(1991)}]{kolmogorov1941b}%
  \BibitemOpen
  \bibfield  {author} {\bibinfo {author} {\bibfnamefont {A.~N.}\ \bibnamefont
  {{Kolmogorov}}},\ }\bibfield  {title} {\bibinfo {title} {{The local structure
  of turbulence in incompressible viscous fluid for very large Reynolds
  numbers}},\ }\href {https://doi.org/10.1098/rspa.1991.0075} {\bibfield
  {journal} {\bibinfo  {journal} {Proceedings of the Royal Society of London
  Series A}\ }\textbf {\bibinfo {volume} {434}},\ \bibinfo {pages} {9}
  (\bibinfo {year} {1991})}\BibitemShut {NoStop}%
\bibitem [{Note8()}]{Note8}%
  \BibitemOpen
  \bibinfo {note} {Note that by \cite {eyink1995}, $|| \protect \overline
  {\protect \bm {\omega }}^{\ell } || \sim \ell ^{-2/3}$ hence eq.~\protect
  \eqref {eq:simil_kolmogorov} scales as $ \ell ^{5/3} \sim k^{-5/3}$ like the
  Kolmogorov formula.}\BibitemShut {Stop}%
\bibitem [{Note9()}]{Note9}%
  \BibitemOpen
  \bibinfo {note} {The \textcolor {black}{approximation} of this calculation
  \textcolor {black}{is based} on equating the Kolmogorov scaling $E^{KOL}(k)$
  with $E(k)$ of eq.~\protect \eqref {eq:simil_kolmogorov} at Kolmogorov
  micro-scale $\eta =(\nu ^3/\langle \varepsilon \rangle )^{1/4}$ for which
  $k_\eta /k_f= 344$. Even though this wavenumber goes beyond the resolution of
  the analysed dataset, fig.~\ref {fig:spec_comp} potentially indicates that
  the two curves do not intersect. From an analytical point of view, \protect
  \emph {large} values of $k$ would make \protect \eqref {eq:kolm_const}
  independent of both the viscosity $\nu $ the (resolved-scale) enstrophy where
  the latter is due to the legitimacy of the approximation $ ||\protect \bm
  {\protect \overline {\omega }}^{\pi /k} || \approx || \protect \bm {{\omega
  }}|| $.}\BibitemShut {Stop}%
\bibitem [{\citenamefont {Donzis}\ and\ \citenamefont
  {Sreenivasan}(2010)}]{donzis2010}%
  \BibitemOpen
  \bibfield  {author} {\bibinfo {author} {\bibfnamefont {D.~A.}\ \bibnamefont
  {Donzis}}\ and\ \bibinfo {author} {\bibfnamefont {K.~R.}\ \bibnamefont
  {Sreenivasan}},\ }\bibfield  {title} {\bibinfo {title} {The bottleneck effect
  and the kolmogorov constant in isotropic turbulence},\ }\href
  {https://doi.org/10.1017/S0022112010001400} {\bibfield  {journal} {\bibinfo
  {journal} {Journal of Fluid Mechanics}\ }\textbf {\bibinfo {volume} {657}},\
  \bibinfo {pages} {171–188} (\bibinfo {year} {2010})}\BibitemShut {NoStop}%
\bibitem [{\citenamefont {Kol{\'a}{\v{r}}}(2007)}]{kolar2007}%
  \BibitemOpen
  \bibfield  {author} {\bibinfo {author} {\bibfnamefont {V.}~\bibnamefont
  {Kol{\'a}{\v{r}}}},\ }\bibfield  {title} {\bibinfo {title} {Vortex
  identification: new requirements and limitations},\ }\href@noop {} {\bibfield
   {journal} {\bibinfo  {journal} {International journal of heat and fluid
  flow}\ }\textbf {\bibinfo {volume} {28}},\ \bibinfo {pages} {638} (\bibinfo
  {year} {2007})}\BibitemShut {NoStop}%
\bibitem [{\citenamefont {Das}\ and\ \citenamefont {Girimaji}(2020)}]{das2020}%
  \BibitemOpen
  \bibfield  {author} {\bibinfo {author} {\bibfnamefont {R.}~\bibnamefont
  {Das}}\ and\ \bibinfo {author} {\bibfnamefont {S.~S.}\ \bibnamefont
  {Girimaji}},\ }\bibfield  {title} {\bibinfo {title} {Revisiting turbulence
  small-scale behavior using velocity gradient triple decomposition},\
  }\href@noop {} {\bibfield  {journal} {\bibinfo  {journal} {New Journal of
  Physics}\ }\textbf {\bibinfo {volume} {22}},\ \bibinfo {pages} {063015}
  (\bibinfo {year} {2020})}\BibitemShut {NoStop}%
\bibitem [{\citenamefont {Eyink}(1995)}]{eyink1995}%
  \BibitemOpen
  \bibfield  {author} {\bibinfo {author} {\bibfnamefont {G.~L.}\ \bibnamefont
  {Eyink}},\ }\bibfield  {title} {\bibinfo {title} {Local energy flux and the
  refined similarity hypothesis},\ }\href@noop {} {\bibfield  {journal}
  {\bibinfo  {journal} {Journal of Statistical Physics}\ }\textbf {\bibinfo
  {volume} {78}},\ \bibinfo {pages} {335} (\bibinfo {year} {1995})}\BibitemShut
  {NoStop}%
\end{thebibliography}%

\end{document}